\documentclass[12pt]{article}
\usepackage{amsmath,siunitx,booktabs}
\usepackage{graphicx,psfrag,epsf}
\usepackage{enumerate}
\usepackage{natbib}
\usepackage{url} 
\usepackage{multirow}
\usepackage{subcaption}
\usepackage{float}
\usepackage[T1]{fontenc}

\newcommand{\blind}{0}

\usepackage{amssymb}

\begin{document}

\def\spacingset#1{\renewcommand{\baselinestretch}%
{#1}\small\normalsize} \spacingset{1}


\if0\blind
{
  \title{\bf Models Based on Exponential Interarrival Times for Single-Unusual-Event Count Data}
  \author{Wanrudee Skulpakdee \\
    Graduate School of Applied Statistics\\
     National Institute of Development Administration\\
     Bangkok, Thailand\\
    and \\
    Mongkol Hunkrajok \\
    Independent Researcher\\
    Bangkok, Thailand\\ }
  \maketitle
} \fi

\if1\blind
{
  \bigskip
  \bigskip
  \bigskip
  \begin{center}
    {\LARGE\bf Models Based on Exponential Interarrival Times for Single-Unusual-Event Count Data}
\end{center}
  \medskip
} \fi

\bigskip
\begin{abstract}
At least one unusual event appears in some count datasets. It will lead to a more concentrated (or dispersed) distribution than the Poisson, the gamma, the Weibull, and the Conway-Maxwell-Poisson (CMP) can accommodate. These well-known count models are based on the equal rates of interarrival times between successive events. Under the assumption of unequal rates (one unusual event) and independent exponential interarrival times, a new class of parametric models for single-unusual-event (SUE) count data is proposed. These two models are applied to two empirical applications, the number of births and the number of bids, and yield considerably better results to the above well-known count models.
\end{abstract}

\noindent%
\vfill

\newpage
\spacingset{1.45} 
\section{Introduction}
\label{sec:intro}
Count data regression analysis is a collection of statistical techniques for modeling and investigating the conditional count distributions of count response variables given sets of covariates. The conditional-variance-mean function of these distributions can be classified into two different categories: linear and non-linear.
\begin{enumerate}
	\item If the distributions are equidispersed (variance $=$ mean), this function is linear.
	\item If the distributions are overdispersed (variance $>$ mean), this function is either linear or non-linear.
	\item If the distributions are underdispersed (variance $<$ mean), this function is either linear or non-linear.
	\item If the distributions are over-, under-, and equidispersed, this function is non-linear.
\end{enumerate}

A renewal process is a counting process.  Its times between successive events are independent and identically distributed with a non-negative distribution (\citealt{Ross10}). The primary assumption of the Poisson model is that the times between events are exponential. It follows that the Poisson model is equidispersed, and the Poisson regression model has a linear conditional-variance-mean function. The exponential distribution replaced by a less restrictive non-negative distribution such as the gamma and the Weibull distributions leads to the gamma (\citealt{Winkelmann95}) and the Weibull (\citealt{McShane08}) count models. They allow for both overdispersion and underdispersion. The gamma and the Weibull regression models have linear conditional-variance-mean functions when the additional parameter ($\alpha$) equals 1, that is, the Poisson regression model. Furthermore, they have nearly linear conditional-variance-mean functions shown in Figures 1(a) and 1(b), although $\alpha$ does not approach 1. 

\begin{figure}[t!]
	\begin{subfigure}{0.5\textwidth}
		\centering
		\includegraphics[width=.9\linewidth]{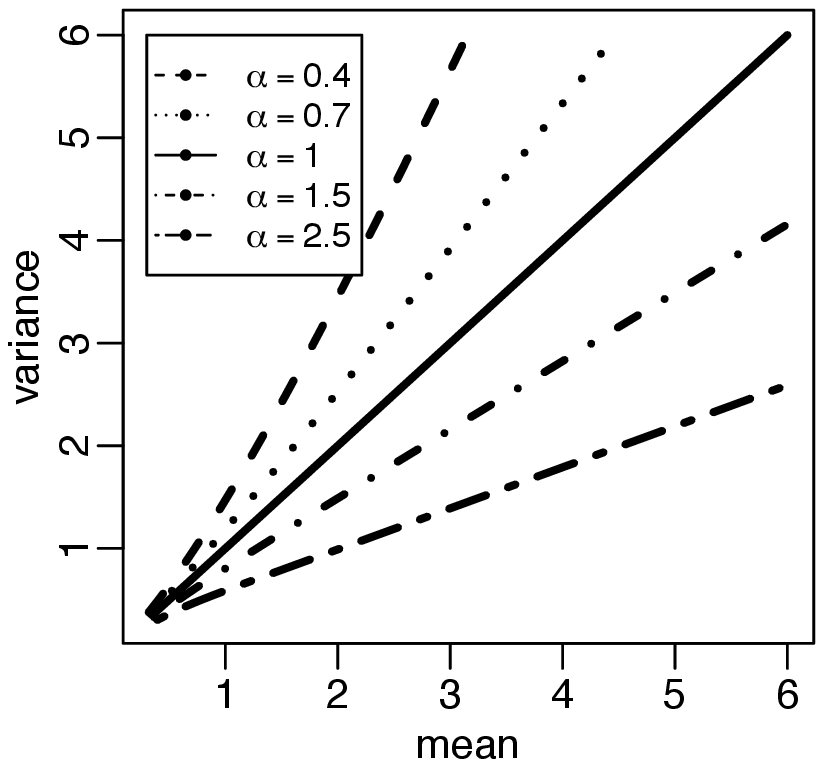}
		\caption{Gamma}
		\label{fig:sfig1}
	\end{subfigure}%
	\begin{subfigure}{0.5\textwidth}
		\centering
		\includegraphics[width=.9\linewidth]{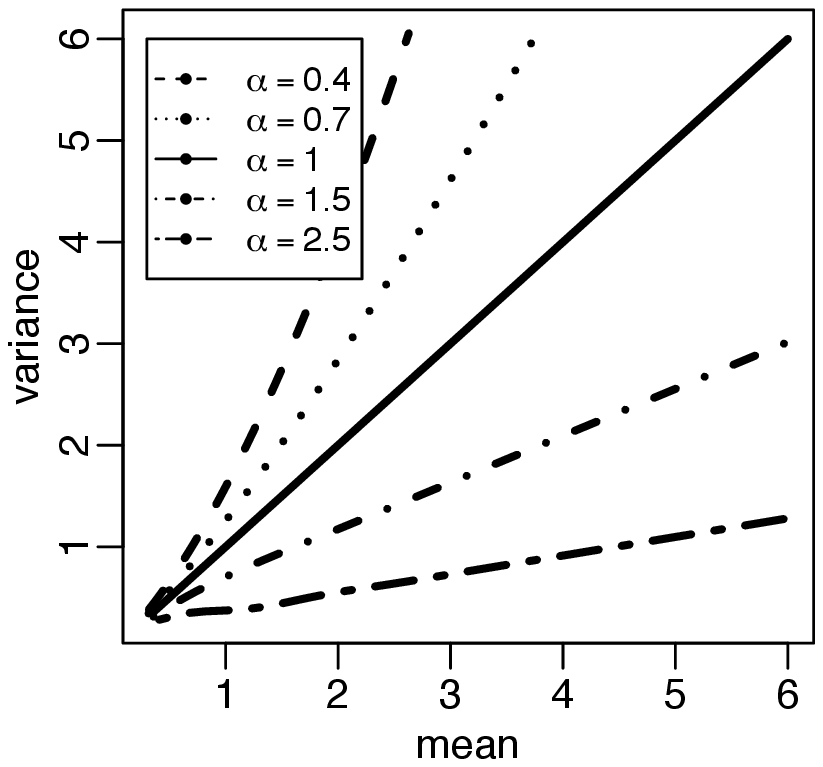}
		\caption{Weibull}
		\label{fig:sfig2}
	\end{subfigure}
	\begin{subfigure}{0.5\textwidth}
		\centering
		\includegraphics[width=.9\linewidth]{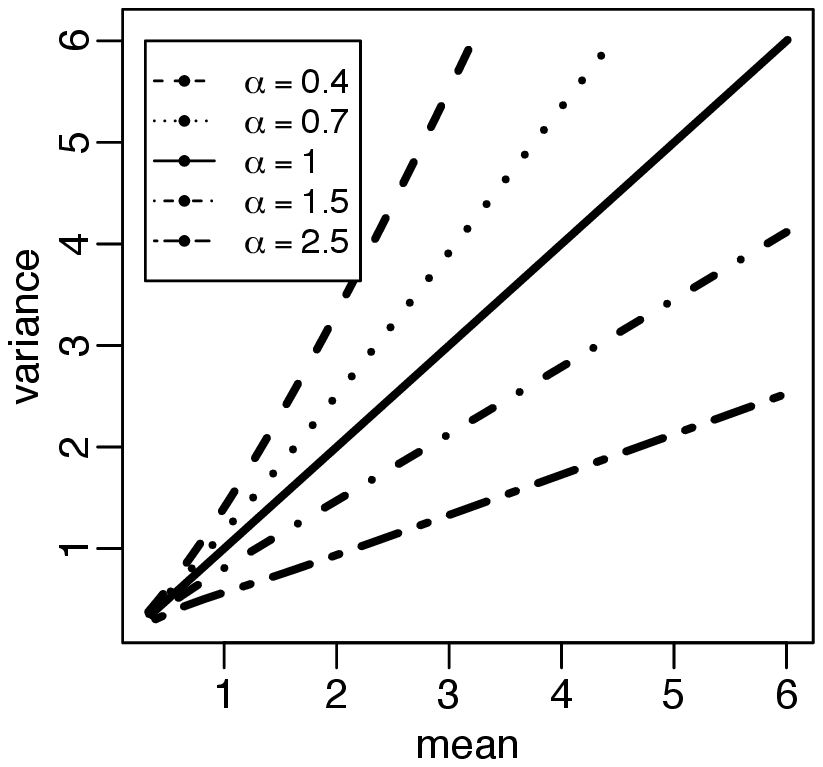}
		\caption{CMP}
		\label{fig:sfig3}
	\end{subfigure}
	\begin{subfigure}{0.5\textwidth}
		\centering
		\includegraphics[width=.9\linewidth]{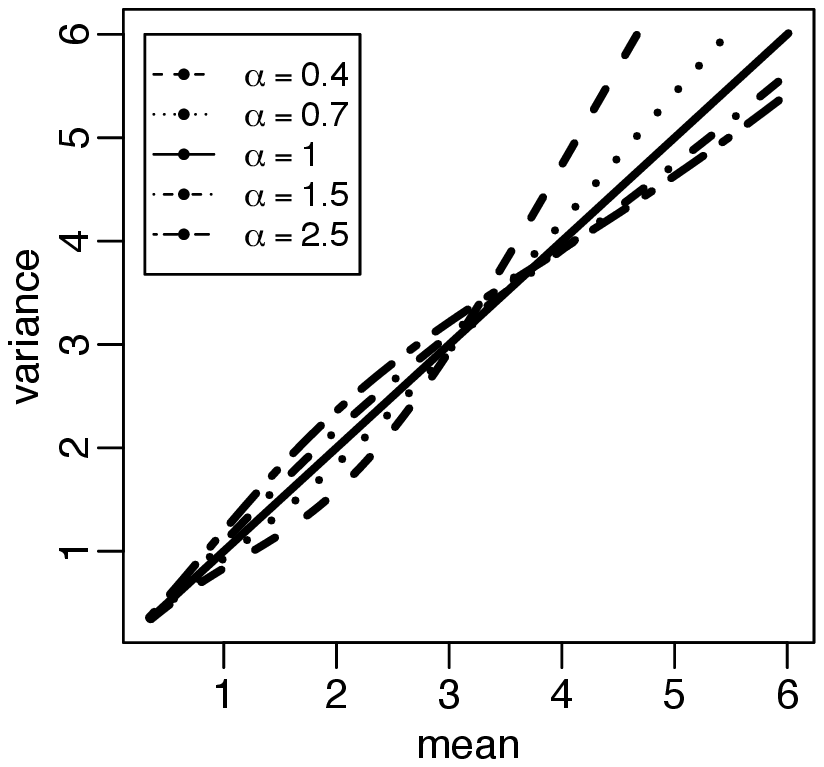}
		\caption{SUE ($\gamma=3$)}
		\label{fig:sfig4}
	\end{subfigure}
	\caption{Graphs showing the linear ($\alpha=1$) and the non-linear ($\alpha \neq 1$) functions of variance and mean}
	\label{fig:fig}
\end{figure}

\sloppy
The Conway-Maxwell-Poisson (CMP) regression model was introduced by \citealt{Kimberly10}. In contrast to the above models, the CMP model is not derived from an underlying renewal process. Surprisingly, however, the graphs in Figures 1(a) and 1(c) of the conditional-variance-mean functions for the gamma and the CMP are hardly distinguishable. A plausible explanation for this similarity is the equality of their approximate variance-mean ratios. These ratios are equal to a constant $1/\alpha$ (\citealt[p.~470]{Winkelmann95}; \citealt[p.~946]{Kimberly10}). Likewise the gamma and the Weibull count models, the CMP model consists of two parameters. Thus, it allows for both overdispersion and underdispersion. Note that the conditional variances and means in Figure 1 were computed in \textbf{R} \citep{RCore} by the \texttt{dCount-conv-bi} function in the \textbf{Countr} package (\citealt{Countr}) for the gamma and the Weibull count models and the \texttt{dcmp} function in the \textbf{COMPoissonReg} package (\citealt{COMPoissonReg}) for the CMP count model.

As previously mentioned, the conditional variance and mean of the above well-known regression models are (nearly) linearly related. In some econometric applications, these regression models are either unsatisfactory or inappropriate when the sample relative frequency distribution is created as a mixture of distributions whose relationship between the variance and the mean is non-linear.

The common assumption that the rates of interarrival times are equal may cause a (nearly) linear conditional-variance-mean function. One potential solution to this problem is to allow the unequal rates. Figure 1(d) shows a single-unusual-event (SUE) count model for five values of $\alpha$. The model has a non-linear conditional-variance-mean function because the rate of the exponential interarrival time between the 2nd and the 3rd differs from others. In Figure 1(d), the 45-degree (Poisson) line indicates that the SUE count model generalizes (nests) the Poisson. The curves corresponding to $ \alpha \neq 1 $ always cross the 45-degree line implies that the count model displays over-, under-, and equidispersion. The development and exploration of the SUE count models are the main objectives of this article.

The rest of this article is organized as follows. Section 2 gives the derivation of a probability function for count models. Some well-known count models are reviewed in Section 3. Section 4 explains how to create the SUE count models and provides two forms of the SUE probability functions required to perform computation. Section 5 describes the SUE regression models. Section 6 provides and analyzes the experimental results from two count datasets, the number of births and the number of bids. Finally, Section 7 concludes the paper.

\section{A Probability Function for Count Models}
Let $X_k, \;k=1,2,...,$ be independent random variables. They are called the interarrival times between the $(k-1)$th and the $k$th event, and their probability density functions $f_{X_k}(x_k)$ are not necessarily identical. The arrival time $S_n$ is the time of the $n$th event. It can be computed by the sum of the interarrival times, $S_n = \sum_{k=1}^{n}X_k$. We will construct a probability function for count models in which an event occurs from time to time. The first event occurs at time $S_1$, the second event at time $S_2$, the third event at time $S_3$, etc. Since $X_k$ are independent, the probability density functions of $S_n$ can be computed by the n-fold convolution of $f_{X_k}(x_k)$,
\begin{align*}
f_{S_n}(s_n) &=\int_{0}^{s_n} \! \!\! \!... \!\int_{0}^{s_2} \! \!
f_{X_1}(s_1)f_{X_2}(s_2 - s_1) ...
f_{X_{n}}(s_{n}-s_{n-1})d{s_1} ... d{s_{n-1}}.\tag{1}
\end{align*}

We note that $S_1=X_1$, and thus $f_{S_1}(s_1) =  f_{X_1}(x_1)$.
The integral of $f_{S_n}(s_n)$ from 0 to $t$ is the cumulative distribution function of $S_n$ evaluated at $t$,
\begin{equation*} 
F_{S_n}(t) = \int_{0}^{t}f_{S_n}(s_n)ds_n.\tag{2}
\end{equation*}

Let $N(t)$ be a discrete random variable, representing the total number of events that occur before or at exactly time $t$. More precisely, $N(t) =0$ if $0\leq t < S_1$, and $N(t) = n$ if $S_n\leq t < S_{n+1}$. 
$F_{S_n}(t)$ and $F_{S_{n+1}}(t)$ are probabilities that at least $n$ and  $n+1$ events occur before or at exactly time $t$. It is easily seen that the probability of the count variable $N(t)$ is given by
\begin{equation*} 
P\left \{ N\left ( t \right )=n \right \}=
\begin{cases}
\: \displaystyle 1-F_{S_{1}}(t)   &   \text{if }   n=0 \\
\: \displaystyle F_{S_n}(t)-F_{S_{n+1}}(t)  &   \text{if } n=1,2,....\tag{3}\\
\end{cases}
\end{equation*}
By substituting (2) into (3) for $n=1,2,...$, we get 
\begin{align*}
P\left \{ N(t)=n \right \} \!&=\! \int_{0}^{t}f_{S_n}\left(s_n\right)ds_n\! -\! \int_{0}^{t}f_{S_{n+1}}\left(s_{n+1}\right)ds_{n+1} \\
&=\! \int_{0}^{t}f_{S_n}(s_n)ds_n\! -\! \int_{0}^{t}\!\!\int_{0}^{s_{n+1}}f_{S_n}(s_n)f_{X_{n+1}}(s_{n+1}\! -\! s_{n}) ds_{n}ds_{n+1}\\
&=\! \int_{0}^{t}f_{S_n}(s_n)ds_n\! -\! \int_{0}^{t}\!\!\int_{s_n}^{t}f_{S_n}(s_n)f_{X_{n+1}}(s_{n+1}\! -\! s_{n})ds_{n+1} ds_{n} \\
&=\! \int_{0}^{t}f_{S_n}(s_n)\left [ 1 -\int_{s_n}^{t}\!f_{X_{n+1}}(s_{n+1}-s_{n})ds_{n+1}\right ]ds_{n} \\
&=\! \int_{0}^{t}f_{S_n}(s_n)\left [ 1 -\int_{0}^{t-s_n}f_{X_{n+1}}(x_{n+1})dx_{n+1} \right ]ds_{n} \\
&=\! \int_{0}^{t}f_{S_n}(s_n)\left [ 1-F_{X_{n+1}}\left(t-s_{n}\right)\right ]ds_{n} \\
&=\! \int_{0}^{t}f_{S_n}(s_n)S_{X_{n+1}}\left(t-s_{n}\right)ds_{n},\tag{4} 
\end{align*}
where $F_{X_{n+1}}(x_{n+1})$ represents the cumulative distribution function of $X_{n+1}$, and  $S_{X_{n+1}}(x_{n+1})$ represents the survival function of  $X_{n+1}$. $S_{X_{n+1}}(t-s_{n}) $ denotes  the probability that the  $(n+1)$th event does not occur after time $S_n$ and before or at exactly time $t$. The probability of $N(t)$ can be described as a convolution of the probability density function of $S_{n}$ and the survival function of $X_{n+1}$. Note that Equation (4) is identical to \citet[p.~424]{Ross10}.

\section{Some Well-known Count Data Models}
If count data are equidispersed or underdispersed, then the Poisson regression model is either unsatisfactory or inappropriate for fitting them. The mixture of conditional Poisson distributions, which are equidispersed, is always overdispersed. The Poisson distribution is equidispersed because its interarrival times $X_k$ are independent and identically distributed with an exponential distribution. Therefore, many count models have been proposed to address this problem.  This limitation may be overcome if the exponential distribution of $X_k$ is assumed to be another distribution. \cite{Winkelmann95} utilized the gamma distribution for the gamma count model,  \cite{McShane08} the Weibull distribution for the Weibull count model, etc. The probability density and the cumulative distribution functions of $X_k$ for the three well-known models are summarized as follows:\\
Poisson count model:
\begin{equation*}
f_{X_k}(x_k) = \lambda e^{-\lambda x_k},\;\; \text{and}\;\; F_{X_k}(x_k) =1-  e^{-\lambda x_k}.\;\;\;\;\;\;                     
\end{equation*}
Gamma count model:
\begin{equation*}
f_{X_k}(x_k) = \frac{\lambda ^\alpha}{\Gamma \left ( \alpha  \right )} x_k^{\alpha-1 }e^{-\lambda x_k},\;\; \text{and}\;\; F_{X_k}(x_k) =\frac{\lambda ^\alpha}{\Gamma \left ( \alpha  \right )}\int_{0}^{x_k} \upsilon ^{\alpha-1}e^{-\lambda\upsilon }d\upsilon.                     
\end{equation*}
Weibull count model:
\begin{equation*}
f_{X_k}(x_k) = \lambda \alpha x_k^{\alpha-1}e^{-\lambda x_k^{\alpha}},\;\; \text{and}\;\; F_{X_k}(x_k) =1- e^{-\lambda x_k^\alpha}.\;\;\;\;\;\;\;\;\;\;\;\;\;\;\;\,                    
\end{equation*}
In this paper, $\lambda$ and $\alpha$ are called the rate and the shape parameters, respectively. The Poisson count model has only the rate parameter, but the last two count models consist of the two parameters. For the Poisson count model, the rate parameter alone determines its distribution entirely. In contrast to the gamma and the Weibull count models, the rate parameters alone determine their distributions partially. The condition of the shape parameter $0<\alpha<1$ corresponds to overdispersion, $\alpha>1$ to underdispersion, and  $\alpha=1$ to equidispersion. In other words, the dispersion types are defined by $\alpha$. The gamma and the Weibull count models generalize the Poisson. It means that when $\alpha=1$, these two count models simply reduce to the Poisson.

Another model that is currently popular for modeling both overdispersed and underdispersed data is the CMP  count model \citep{Kimberly10}. This two-parameter ($\lambda$ and $\alpha$) model generalizes the Poisson, but the distributions of the interarrival times are not identical. It might not be easy to derive the probability density and the cumulative distribution functions of $X_k$, except for $f_{X_1}\left ( x_1 \right )=\frac{1}{z\left ( \lambda x_1,\alpha  \right )^2}\frac{\mathrm{d}z\left ( \lambda x_1,\alpha  \right )}{\mathrm{d} x_1}$ and $F_{X_1}\left ( x_1 \right )=1- \frac{1}{z\left ( \lambda x_1,\alpha \right )}$, where $z\left ( \lambda x_1,\alpha  \right )=\sum_{s=0}^{\infty }\frac{(\lambda x_1) ^s}{\left ( s! \right )^\alpha }$. Note that the dispersion types defined by $\alpha$ are identical to the gamma and the Weibull count models.

\section{SUE Count Models}
Count data are the number of events in a given interval of time. The Poisson, the gamma, the Weibull, and the CMP cannot handle some count datasets with different desired events because their distributions are less concentrated (or dispersed) than the data distributions. In other words, there is at least one unusual event in these datasets. For example, the number of births by a woman was described and analyzed by \citet{Winkelmann95}, \citet{McShane08}, and \citet{Kharrat19}. Figure 4(a) illustrates the sample probabilities of this dataset. Since the sample probability of the two children is much greater than the other numbers, one question is, do these women want to have all children the same? The experiments in Section 6 reveal that the women need a third child less than other children. Therefore, having a third child is an unusual event.

In the context of a counting process, events $k \in \boldsymbol{\gamma}=\left \{ \gamma _{1}, \gamma _{2},...,\gamma _{d}\right \}$ are unusual if $X_k$ have arbitrary distributions with rates $\lambda_{k\in \boldsymbol{\gamma}}\neq\lambda$ and $\lambda_{k\notin \boldsymbol{\gamma}}=\lambda$. The possible distributions are the exponential, the gamma, the Weibull, etc. The dynamic hurdle Poisson model was introduced by \citet{Baetschmann17}. This model is motivated by a desire to explain excess zeros in the count data. Thus, it is assumed that 
\begin{equation*} 
f_{X _{k}}\left (x_k\right )=
\begin{cases}
\: \lambda_1 e^{-\lambda_1 x_1}   &    \text{if }  k=1 \\
\: \lambda_2 e^{-\lambda_2 x_k}    &   \text{if }   k\neq 1,   \\
\end{cases}
\end{equation*}
and
\begin{equation*} 
F_{X_{k}}\left (x_k\right )=
\begin{cases}
\: 1-e^{-\lambda_1 x_1}   &    \text{if }  k=1\\
\: 1-e^{-\lambda_2 x_k}    &   \text{if }   k\neq 1.\\
\end{cases}
\end{equation*}\\
It is easily seen that the first event is unusual, and $X_k$ have exponential distributions. For the dynamic hurdle Poisson regression model, the two parameters $\lambda_1$ and $\lambda_2$ are functions of covariates. Consequently, the number of its estimated parameters is twice the Poisson and approximately twice the gamma, the Webull, and the CMP. This regression model can be adapted to fit a count dataset whose unusual event is not the first event, but its double parameters comparing to the above well-known count models may lead to the risk of overfitting. Perhaps, satisfactory or appropriate count models are described below.

The count models whose $X_k$ are independent and not identically distributed will be proposed. In this research, the models are assumed to have only a single unusual event. A simple way of constructing the single-unusual-event (SUE) count models is to assume that $X_k$ are exponential with an unusual rate $\lambda_{k=\gamma}=\alpha\lambda$ and usual rates $\lambda_{k\neq\gamma}=\lambda$. More precisely,
\begin{equation*} 
f_{X _{k}}\left (x_k\right )=
\begin{cases}
\: \alpha \lambda e^{-\alpha \lambda x_k}   &    \text{if }  k=\gamma \\
\: \lambda e^{-\lambda x_k}    &   \text{if }   k\neq \gamma,\tag{5}   \\
\end{cases}
\end{equation*}
and
\begin{equation*} 
F_{X_{k}}\left (x_k\right )=
\begin{cases}
\: 1- e^{-\alpha \lambda x_k}   &    \text{if }  k=\gamma\\
\: 1-e^{-\lambda x_k}    &   \text{if }   k\neq \gamma.\tag{6}   \\
\end{cases}
\end{equation*}\\
Using Equations (3)-(6), the SUE count models are obtained as follows:\\
If $\gamma=1$,
\begin{equation*} 
P\left \{ N\left ( t \right )=n \right \}=
\begin{cases}
\: \displaystyle e^{-\alpha \lambda t}   &   \text{if }   n=0 \\
\: \displaystyle \frac{\alpha e^{-\lambda t}}{\left ( 1-\alpha  \right )^{n}}\left ( e^{\left ( 1-\alpha \right )\lambda t}-\sum\limits_{i=0}^{n-1}\frac{\left (\left ( 1-\alpha     \right )\lambda t\right )^{i}}{i!} \right) &    \text{if }  n>0. \tag{7}\\
\end{cases}
\end{equation*}
If $\gamma>1$,
\begin{equation*} 
P\left \{ N\left ( t \right )=n \right \}=
\begin{cases}
\: \displaystyle \frac{\left ( \lambda t \right )^{n}e^{-\lambda t}}{n!}  &   \text{if }   n< \gamma -1 \\
\: \displaystyle \frac{ e^{-\lambda t}}{\left ( 1-\alpha  \right )^{n}}\left ( e^{\left ( 1-\alpha \right )\lambda t}-\sum\limits_{i=0}^{n-1}\frac{\left (\left ( 1-\alpha     \right )\lambda t\right )^{i}}{i!} \right)  &    \text{if }  n=\gamma -1 \\
\: \displaystyle \frac{\alpha e^{-\lambda t}}{\left ( 1-\alpha  \right )^{n}}\left ( e^{\left ( 1-\alpha \right )\lambda t}-\sum\limits_{i=0}^{n-1}\frac{\left (\left ( 1-\alpha     \right )\lambda t\right )^{i}}{i!} \right) &    \text{if }  n>\gamma -1. \tag{8}\\
\end{cases}
\end{equation*}\\
For $n<\gamma -1$, the probability function is Poisson. For $n=\gamma -1$, the probability function $\frac{ e^{-\lambda t}}{\left ( 1-\alpha  \right )^{n}}\left ( e^{\left ( 1-\alpha \right )\lambda t}-\sum_{i=0}^{n-1}\frac{\left (\left ( 1-\alpha     \right )\lambda t\right )^{i}}{i!} \right)$ can be simplified to the sum of a Poisson function and an infinite series of functions $\frac{ \left ( \lambda t \right )^{n}e^{-\lambda t}}{n!}+\frac{ e^{-\lambda t}}{\left ( 1-\alpha  \right )^{n}}\sum_{i=n+1}^{\infty}\frac{\left (\left ( 1-\alpha     \right )\lambda t\right )^{i}}{i!}$.  Since the second term is positive ($\alpha<1$) and negative ($\alpha>1$), the SUE probability value is greater and smaller than the Poisson ($\alpha=1$), respectively (see Figure 2).

Additionally, the first and the second moments of the SUE probability distributions in closed or finite form, which are the expected values of $N(t)$ and $N(t)^2$, are given as follows:\\
If $\gamma=1$,
\begin{equation*} 
E\left \{ N\left ( t \right ) \right \}=\lambda t+\left ( \frac{\alpha-1}{\alpha } \right )\left ( 1-e^{-\alpha \lambda t} \right ),
\end{equation*}
and
\begin{equation*} 
E\left \{ N\left ( t \right )^{2} \right \}=\left ( \frac{3\alpha -2}{\alpha }\right )\lambda t +\left ( \lambda t \right )^{2}+\frac{\left (\alpha- 2 \right )\left ( \alpha-1  \right )}{\alpha ^{2}}\left ( 1-e^{-\alpha \lambda t} \right ).
\end{equation*}
If $\gamma>1$,
\begin{equation*}
\begin{split}
E\left \{ N\left ( t \right ) \right \}=&\sum_{n=1}^{\gamma -2}\frac{n\left ( \lambda t \right )^{n}e^{-\lambda t}}{n!}-\sum_{n=1}^{\gamma -2}\frac{n\alpha e^{-\lambda t}}{\left ( 1-\alpha  \right )^{n}}\left ( e^{\left ( 1-\alpha  \right )\lambda t}-\sum_{i=0}^{n-1}\frac{\left ( \left ( 1-\alpha  \right )\lambda t \right )^{i}}{i!} \right )\\
&+\frac{\left ( \gamma -1 \right )e^{-\lambda t}}{\left ( 1-\alpha  \right )^{\gamma -2}}\left ( e^{\left ( 1-\alpha  \right )\lambda t}-\sum_{i=0}^{\gamma -2}\frac{\left ( \left ( 1-\alpha  \right )\lambda t \right )^{i}}{i!} \right ) \\
&+\lambda t+\left ( \frac{\alpha-1}{\alpha } \right )\left ( 1-e^{-\alpha \lambda t} \right ),
\end{split}
\end{equation*}
and
\begin{equation*}
\begin{split}
E\left \{ N\left ( t \right )^{2} \right \}=&\sum_{n=1}^{\gamma -2}\frac{n^{2}\left ( \lambda t \right )^{n}e^{-\lambda t}}{n!}-\sum_{n=1}^{\gamma -2}\frac{n^{2}\alpha e^{-\lambda t}}{\left ( 1-\alpha  \right )^{n}}\left ( e^{\left ( 1-\alpha  \right )\lambda t}-\sum_{i=0}^{n-1}\frac{\left ( \left ( 1-\alpha  \right )\lambda t \right )^{i}}{i!} \right )\\
&+\frac{\left ( \gamma -1 \right )^{2}e^{-\lambda t}}{\left ( 1-\alpha  \right )^{\gamma -2}}\left ( e^{\left ( 1-\alpha  \right )\lambda t}-\sum_{i=0}^{\gamma -2}\frac{\left ( \left ( 1-\alpha  \right )\lambda t \right )^{i}}{i!} \right ) \\
&+\left ( \frac{3\alpha -2}{\alpha }\right )\lambda t +\left ( \lambda t \right )^{2}+\frac{\left ( \alpha-2  \right )\left ( \alpha-1  \right )}{\alpha ^{2}}\left ( 1-e^{-\alpha \lambda t} \right ).
\end{split}
\end{equation*}\\

\begin{figure}[H]
	\centering
	\includegraphics[width=0.8\linewidth]{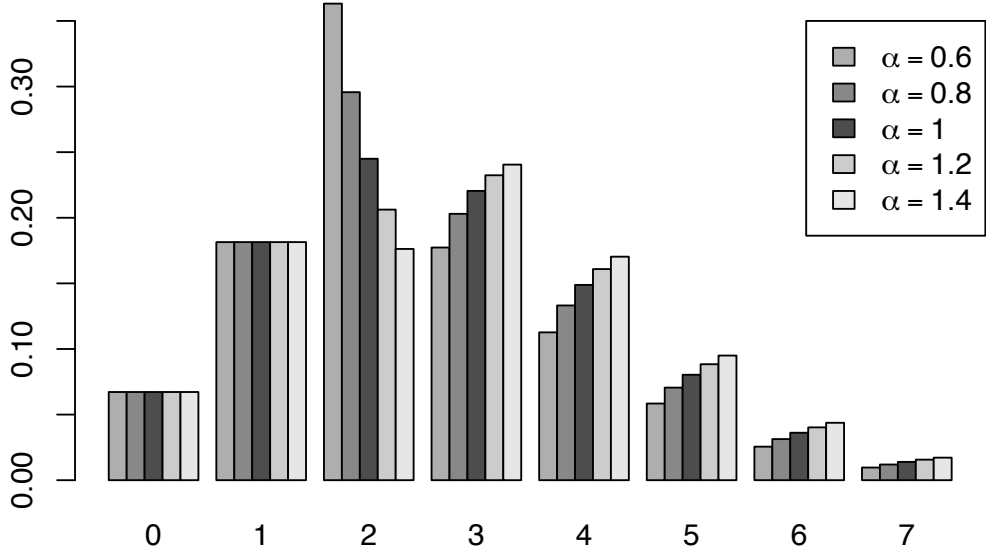}
	\caption{SUE ($\gamma=3$ and $\lambda = 2.7$) distributions with unequal means and dispersions.}
\end{figure}

The  variance of $N\left ( t \right )$ is given by $ V\left \{ N\left ( t \right ) \right \}=E\left \{ N\left ( t \right )^{2} \right \}-E\left \{ N\left ( t \right ) \right \}^2$. The values of 
${\frac{ e^{-\lambda t}}{\left ( 1-\alpha  \right )^{n}}\left ( e^{\left ( 1-\alpha \right )\lambda t}-\sum_{i=0}^{n-1}\frac{\left (\left ( 1-\alpha     \right )\lambda t\right )^{i}}{i!} \right)}$ 
and 
${\frac{ \alpha e^{-\lambda t}}{\left ( 1-\alpha  \right )^{n}}\left ( e^{\left ( 1-\alpha \right )\lambda t}-\sum_{i=0}^{n-1}\frac{\left (\left ( 1-\alpha \right )\lambda t\right )^{i}}{i!} \right)}$ 
in (7) and (8) must be between 0 and 1, but these computed values may not be in this interval. These functions have an indeterminate form $0/0$ when $\alpha=1$.  Furthermore, they may be required for $\alpha$ near 1. $e^{\left (1-\alpha\right )\lambda t }$ is almost equal to $ \sum_{i=0}^{n-1}\frac{\left (\left (1-\alpha\right )\lambda t\right)^{i}}{i!}$. It is easily seen that there is a potential loss of significance in the subtraction, which leads to a possibility of computation inaccuracy during a maximum likelihood estimation.  These two problems can be eliminated by using a Taylor series expansion of $e^{\left ( 1-\alpha  \right )\lambda t}=\sum_{i=0}^{\infty}\frac{\left (\left (1-\alpha\right )\lambda t\right)^{i}}{i!}$. The SUE probability distributions and their first and second moments can be rewritten as follows:\\
If $\gamma=1$,
\begin{equation*} 
P\left \{ N\left ( t \right )=n \right \}=
\begin{cases}
\: \displaystyle e^{-\alpha \lambda t}   &   \text{if }   n=0 \\
\: \displaystyle \alpha \left ( \lambda t \right )^{n}e^{-\lambda t}\sum\limits_{i=0}^{\infty}\frac{\left (\left ( 1-\alpha     \right )\lambda t\right )^{i}}{\left (  i+n\right )!} &    \text{if }  n>0,\tag{9} \\
\end{cases}\\
\end{equation*}
\begin{equation*} 
E\left \{ N\left ( t \right ) \right \}=\lambda t+\left ( \frac{\alpha-1}{\alpha } \right )\left ( 1-e^{-\alpha \lambda t} \right ),\\
\end{equation*}
and
\begin{equation*} 
E\left \{ N\left ( t \right )^{2} \right \}=\left ( \frac{3\alpha -2}{\alpha }\right )\lambda t +\left ( \lambda t \right )^{2}+\frac{\left ( \alpha-2  \right )\left ( \alpha-1 \right )}{\alpha ^{2}}\left ( 1-e^{-\alpha \lambda t} \right ).
\end{equation*}
If $\gamma>1$,
\begin{equation*}
P\left \{ N\left ( t \right )=n \right \}=
\begin{cases}
\: \displaystyle \frac{\left ( \lambda t \right )^{n}e^{-\lambda t}}{n!}  &   \text{if }   n< \gamma -1 \\
\: \displaystyle \left ( \lambda t \right )^{n}e^{-\lambda t}\sum\limits_{i=0}^{\infty}\frac{\left (\left ( 1-\alpha     \right )\lambda t\right )^{i}}{\left (  i+n\right )!} 
&    \text{if }  n=\gamma -1 \\
\: \displaystyle \alpha \left ( \lambda t \right )^{n}e^{-\lambda t}\sum\limits_{i=0}^{\infty}\frac{\left (\left ( 1-\alpha     \right )\lambda t\right )^{i}}{\left (  i+n\right )!} &    \text{if }  n>\gamma -1 ,\tag{10}\\
\end{cases}
\end{equation*}
\begin{equation*}
\begin{split}
&E\left \{ N\left ( t \right ) \right \}=\sum_{n=1}^{\gamma-2 }\frac{n\left ( \lambda t \right )^{n}e^{-\lambda t}}{n!}-\sum_{n=1}^{\gamma -2}n\alpha\left ( \lambda t \right )^{n}e^{-\lambda t} \sum_{i=0}^{\infty }\frac{\left ( \left ( 1-\alpha  \right )\lambda t \right )^{i}}{\left ( i+n \right )!}\\
&\quad\quad\quad\quad\quad+\left ( \gamma -1 \right )\left ( 1-\alpha  \right )\left ( \lambda t \right )^{\gamma -1}e^{-\lambda t}\sum_{i=0}^{\infty }\frac{\left ( \left ( 1-\alpha  \right )\lambda t\right )^{i}}{\left ( i+\gamma -1 \right )!} \\
&\quad\quad\quad\quad\quad+\lambda t+\left ( \frac{\alpha-1}{\alpha } \right )\left ( 1-e^{-\alpha \lambda t} \right ),
\end{split}
\end{equation*}
and
\begin{equation*}
\begin{split}
&E\left \{ N\left ( t \right )^{2} \right \}=\sum_{n=1}^{\gamma-2 }\frac{n^{2}\left ( \lambda t \right )^{n}e^{-\lambda t}}{n!}-\sum_{n=1}^{\gamma -2}n^{2}\alpha\left ( \lambda t \right )^{n}e^{-\lambda t} \sum_{i=0}^{\infty }\frac{\left ( \left ( 1-\alpha  \right )\lambda t \right )^{i}}{\left ( i+n \right )!}\\
&\quad\quad\quad\quad\quad+\left ( \gamma -1 \right )^{2}\left ( 1-\alpha  \right )\left ( \lambda t \right )^{\gamma -1}e^{-\lambda t}\sum_{i=0}^{\infty }\frac{\left ( \left ( 1-\alpha  \right )\lambda t\right )^{i}}{\left ( i+\gamma -1 \right )!} \\
&\quad\quad\quad\quad\quad+\left ( \frac{3\alpha -2}{\alpha }\right )\lambda t +\left ( \lambda t \right )^{2}+\frac{\left ( \alpha-2  \right )\left ( \alpha-1  \right )}{\alpha ^{2}}\left ( 1-e^{-\alpha \lambda t} \right ).
\end{split}
\end{equation*}

Figure 2 compares SUE ($\gamma=3$ and $\lambda = 2.7$) distributions for five values of $\alpha$. For $\alpha=1$, the SUE distribution is a Poisson distribution. For $\alpha < 1$, the smaller $\alpha$ is, the greater the probability value of the two outcomes is, and the more concentrated the distribution becomes. For $\alpha > 1$, the greater $\alpha$ is, the smaller the probability value of the two outcomes is, and the more dispersed the distribution becomes.

\begin{figure}[!t]
	\begin{subfigure}{0.5\textwidth}
		\centering
		\includegraphics[width=.7\linewidth]{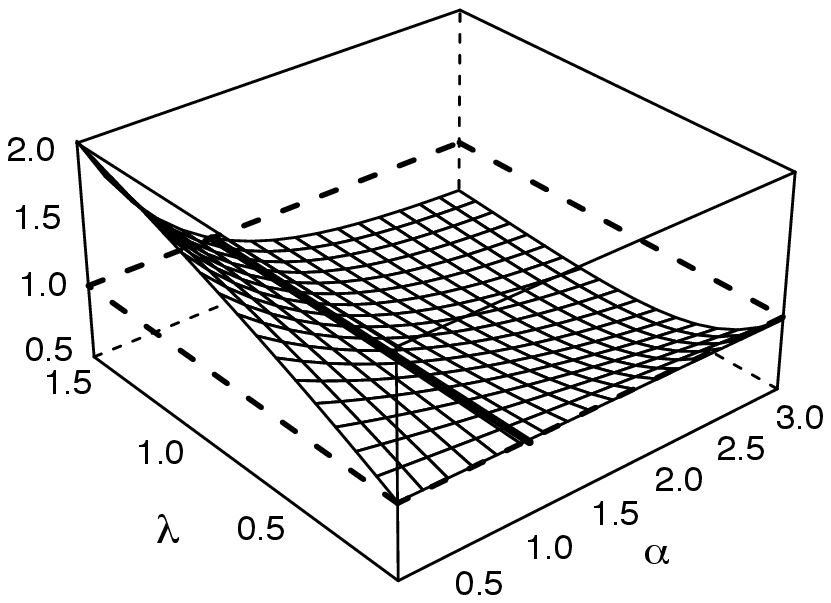}
		\caption{$\gamma =1$}
		\label{fig:sfig1}
	\end{subfigure}%
	\begin{subfigure}{0.5\textwidth}
		\centering
		\includegraphics[width=.7\linewidth]{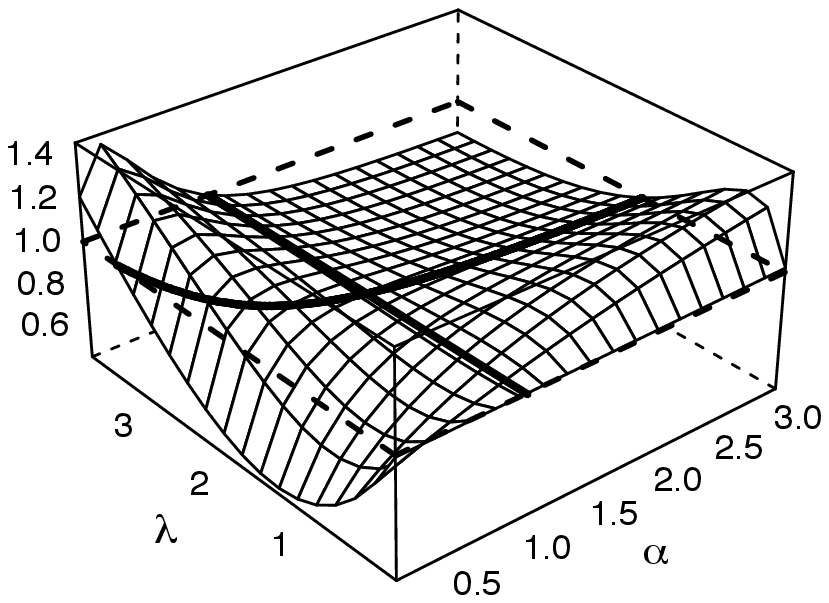}
		\caption{$\gamma =2$}
		\label{fig:sfig2}
	\end{subfigure}
	\begin{subfigure}{0.5\textwidth}
		\centering
		\includegraphics[width=.7\linewidth]{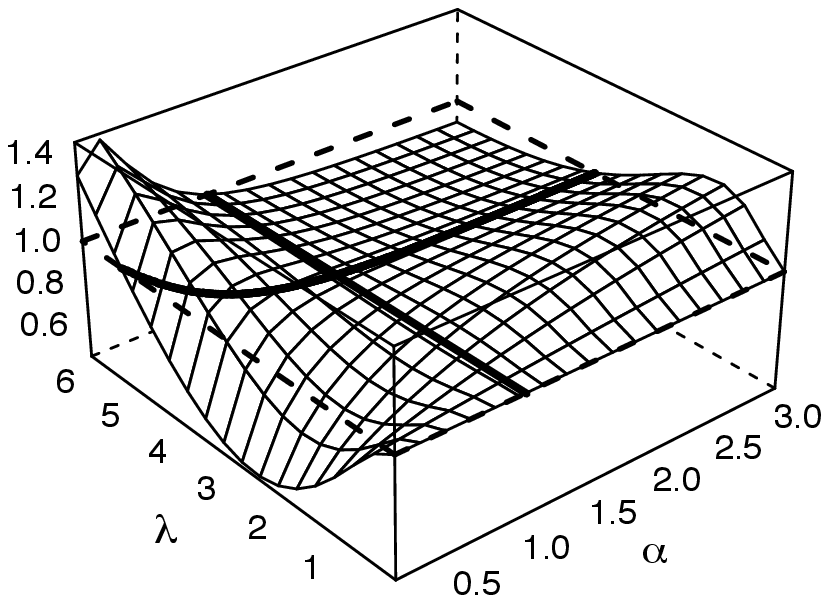}
		\caption{$\gamma =3$}
		\label{fig:sfig3}
	\end{subfigure}
	\begin{subfigure}{0.5\textwidth}
		\centering
		\includegraphics[width=.7\linewidth]{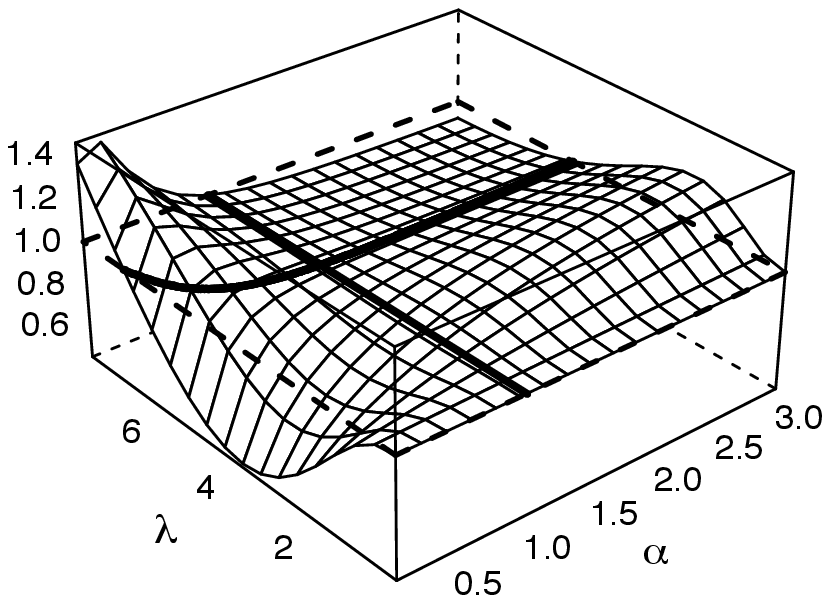}
		\caption{$\gamma =4$}
		\label{fig:sfig4}
	\end{subfigure}
	\begin{subfigure}{0.5\textwidth}
		\centering
		\includegraphics[width=.7\linewidth]{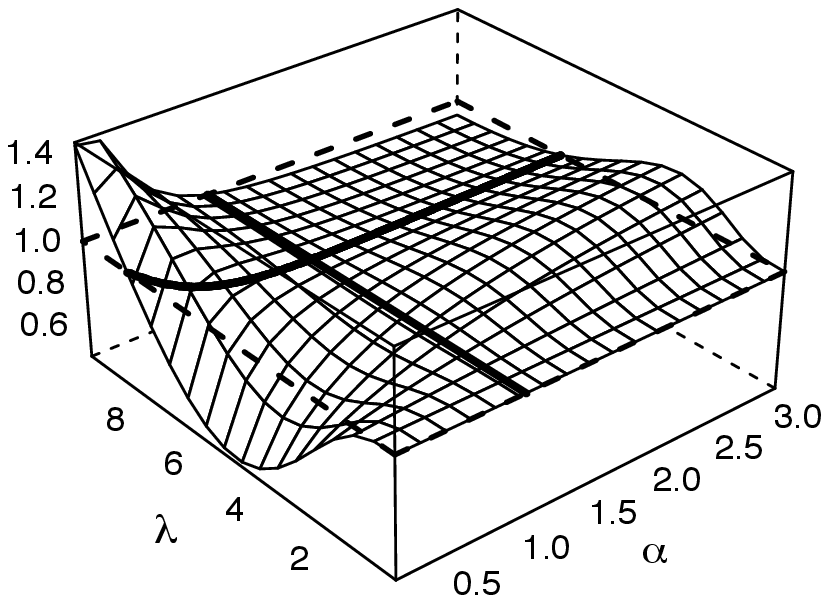}
		\caption{$\gamma =5$}
		\label{fig:sfig5}
	\end{subfigure}
	\begin{subfigure}{0.5\textwidth}
		\centering
		\includegraphics[width=.7\linewidth]{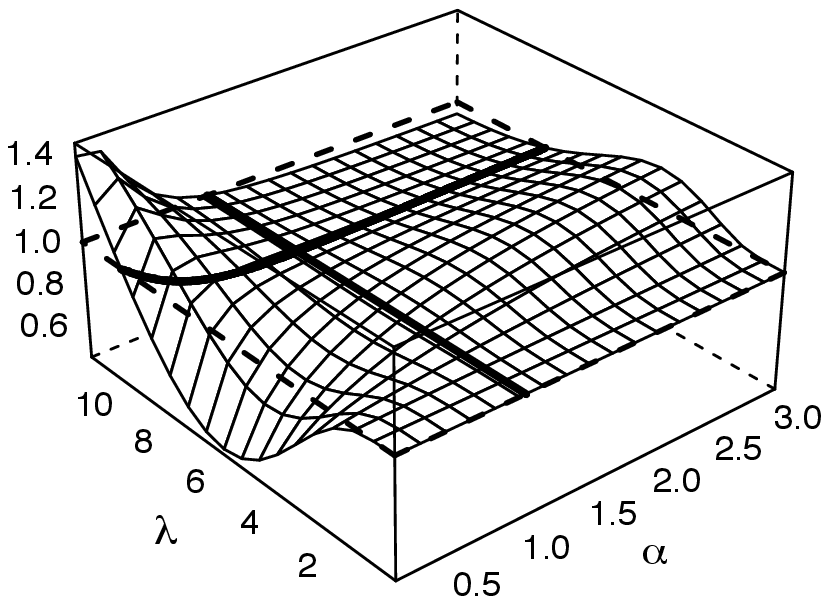}
		\caption{$\gamma =6$}
		\label{fig:sfig6}
	\end{subfigure}
	\caption{Variance-mean ratios for SUE count models with $0<\alpha < 3$. Each surface ($\gamma>1$) contains a saddle point, which is the intersection of the straight and the curved lines.}
	\label{fig:fig}
\end{figure}

Figure 3 shows graphs of the variance-mean ratio with $\gamma=1-6$ for various values of $\lambda$ and $\alpha$. The straight and the curved solid lines in which the $\alpha \lambda$-plane intersects the surface display equidispersion, indicating that the SUE ($\gamma>1$) models can be equidispersed even though  $\alpha$ is not necessarily equal to 1. The values of the variance-mean ratio above and below $\alpha \lambda$-planes represent overdispersion and underdispersion. The gamma, the Weibull, and the CMP count models exhibit overdispersion($0<\alpha<1$), underdispersion($\alpha>1$), and equidispersion ($\alpha=1$). Likewise, the SUE has these dispersion properties but only when $\gamma=1$ (see Figure 3(a)). The proof is shown by subtracting the variance and the mean, that is, $V\left \{ N\left ( t \right ) \right \}-E\left \{ N\left ( t \right ) \right \}$ $=$ $2\left( 1-\alpha  \right )\alpha ^{-2} e^{-\alpha \lambda t}\left ( \alpha \left ( \textrm{cosh}\left ( \alpha \lambda t \right ) -1\right )+\textrm{sinh}\left ( \alpha \lambda t \right )-\alpha \lambda t\right )$. Since $ \textrm{cosh}\left ( \alpha \lambda t \right ) -1$ and $\textrm{sinh}\left ( \alpha \lambda t \right )$ $-\alpha \lambda t$ are always positive, the variance-mean subtraction is positive ($0<\alpha<1$), negative ($\alpha>1$), and zero ($\alpha=1$). For $\gamma>1$, these dispersion properties do not hold (see Figures 3(b)-3(f)). The dispersion types of the SUE ($\gamma>1$) are defined by both $\alpha$ and $\lambda$. However, $\alpha=1$ still indicates equidispersion because the SUE count models generalize the Poisson.

\section{SUE Regression Models}
The outcome or response variables $n_1,...,n_m$ are assumed to be observed values of independent discrete random variables $N_1(t),...,N_m(t)$ such that $N_j(t)$ has the SUE distribution with parameters $\lambda_{j}$ and $\alpha_{j}$. $\lambda_{j}$ and $\alpha_{j}, j=1,...,m,$ are the $j$th observation of the rate and the shape parameters, respectively. The link functions relate the rate and the shape parameters of the SUE distribution to linear predictors, that is, $g_\lambda \left ( \lambda _{j} \right )=\beta_{0}+\beta_{1}x_{j1}+...+\beta_{r}x_{jr}$ and $g_\alpha\left (\alpha_j \right )=\beta_{r+1}$. $x_{jk},k=1,...,r$, is the $j$th observation of the $k$th regressor variable or covariate, and $\beta _{l}, l=0,...,r+1,$ is the $l$th unknown parameter to be estimated. Note that $g_\alpha\left (\alpha_j \right )$ can be defined as a function of the covariates. The inverses of the link functions for the SUE distribution are $\lambda_{j}=e^{\beta_{0}+\beta_{1}x_{j1}+...+\beta_{r}x_{jr}}$ and 
 $\alpha_j=e^{\beta_{r+1}}$, ensuring that $\lambda_{j}$ and $\alpha_{j}$ remain positive for all unknown parameter and covariate combinations. Thus, the link functions are logarithmic functions. The SUE count regression models are given as follows:  
 If $\gamma=1$,
 \begin{equation*} 
 P\left \{ N_j\left( t\right )=n_j\right \}=
 \begin{cases}
 \: \displaystyle e^{-\alpha_j\lambda_j t}   &   \text{if }   n_j=0 \\
 \: \displaystyle \frac{\alpha_j e^{-\lambda_j t}}{\left ( 1-\alpha_j  \right )^{n_j}}\left ( e^{\left ( 1-\alpha_j \right )\lambda_j t}-\sum\limits_{i=0}^{n_j-1}\frac{\left (\left ( 1-\alpha_j\right )\lambda_j t\right )^{i}}{i!} \right) &    \text{if }  n_j>0, \tag{11}\\
 \end{cases}\\
\end{equation*}
and
\begin{equation*}
 P\left \{ N_j\left ( t \right )=n_j \right \}=
 \begin{cases}
 \: \displaystyle e^{-\alpha_j \lambda_j t}   & \quad\quad\quad\quad\quad  \text{if } 
   n_j=0 \\
 \: \displaystyle \alpha_j \left ( \lambda_j t \right )^{n_j}e^{-\lambda_j t}\sum\limits_{i=0}^{\infty}\frac{\left (\left ( 1-\alpha_j \right )\lambda_j t\right )^{i}}{\left (  i+n_j\right )!} & \quad\quad\quad\quad\quad    \text{if }  n_j>0.\tag{12} \\
 \end{cases}
\end{equation*}
If $\gamma>1$,
\begin{equation*}
 P\left \{ N_j\left( t \right )=n_j\right \}=
 \begin{cases}
 \: \displaystyle \frac{\left ( \lambda_j t \right )^{n_j}e^{-\lambda_j t}}{n_j!}  &   \text{if }   n_j< \gamma -1 \\
 \: \displaystyle \frac{ e^{-\lambda_j t}}{\left ( 1-\alpha_j  \right )^{n_j}}\left ( e^{\left ( 1-\alpha_j \right )\lambda_j t}-\sum\limits_{i=0}^{n_j-1}\frac{\left (\left ( 1-\alpha_j\right )\lambda_j t\right )^{i}}{i!} \right)&    \text{if }  n_j=\gamma -1 \\
 \: \displaystyle \frac{\alpha_j e^{-\lambda_j t}}{\left ( 1-\alpha_j  \right )^{n_j}}\left ( e^{\left ( 1-\alpha_j \right )\lambda_j t}-\sum\limits_{i=0}^{n_j-1}\frac{\left (\left ( 1-\alpha_j\right )\lambda_j t\right )^{i}}{i!} \right) &    \text{if }  n_j>\gamma -1, \tag{13}
 \end{cases}\\
\end{equation*}
and
\begin{equation*}
 P\left \{ N_j\left ( t \right )=n_j \right \}=
 \begin{cases}
 \: \displaystyle \frac{\left (\lambda_j t \right )^{n_j}e^{-\lambda_j t}}{n_j!}  &   \quad\quad\quad\quad\quad \text{if }   n_j< \gamma -1 \\
 \: \displaystyle \left (\lambda_j t \right )^{n_j}e^{-\lambda_j t}\sum\limits_{i=0}^{\infty}\frac{\left(\left( 1-\alpha_j\right )\lambda_j t\right )^{i}}{\left (  i+n_j\right )!} 
 &  \quad\quad\quad\quad\quad   \text{if }  n_j=\gamma -1 \\
 \: \displaystyle \alpha_j \left ( \lambda_j t \right )^{n_j}e^{-\lambda_j t}\sum\limits_{i=0}^{\infty}\frac{\left (\left ( 1-\alpha_j\right )\lambda_j t\right )^{i}}{\left (  i+n_j\right )!} &  \quad\quad\quad\quad\quad   \text{if }  n_j>\gamma -1.\tag{14}\\
 \end{cases}\\
\end{equation*}
 
To estimate the unknown parameters $\left(\beta_0,...,\beta_{r+1}\right)$, the log-likelihood function, which is given by
\begin{equation*} 
\ell\left ( \beta_0,...,\beta_{r+1}\right)=\sum_{j=1}^{m}\ln P\left \{ N_j\left ( t \right )=n_j \right \}, \tag{15}
\end{equation*}
is maximized. In the following applications, the optimization methods such as "BFGS", "nlminb" and "CG" were used to find the values of the unknown parameters that maximize the log-likelihood function.

\section{Experimental Results}
This section shows results from two applications. The fertility data were analyzed by \citet{Winkelmann95} and re-analyzed by \citet{McShane08} and \citet{Kharrat19}. The takeover bids data were analyzed by \citet{Jaggia93} and re-analyzed by \citet{Colin97} and \citet{Castillo13}. For more information, the readers are referred to \citet{Winkelmann95} for the fertility data and \citet{Colin97} for the takeover bids data.

Experimental results obtained from the Poisson, the gamma, the Weibull, and the CMP models are computed using the \textbf{stats} \citep{RCore} , \textbf{Countr} (gamma and Weibull) \citep{Countr}, and \textbf{COMPoissonReg} \citep{COMPoissonReg} \textbf{R} packages. The fertility and the takeover bids datasets are available from the \textbf{Countr} and the \textbf{mpcmp} \citep{mpcmp} \textbf{R} packages, respectively. The SUE is implemented in \textbf{R} \citep{RCore} and C++. Most of the code is written in C++ via the \textbf{Rcpp} \citep{Rcpp} package needs for accelerating computations. 

The fertility data, which consists of 10 covariates, are very slightly underdispersed with the variance-mean ratio of 2.328/2.384 = 0.977. 
The Poisson regression model is inappropriate because the mixture of conditional equidispersed distributions is always overdispersed. The gamma, the Weibull, and the CMP models display underdispersion ($\alpha>1$). These regression models perhaps provide a good fit for the data because the mixture of conditional underdispersed distributions can be  over-, under-, or equidispersion. The log-likelihood values computed using the Poisson, the gamma, the Weibull, the CMP, and the SUE ($\gamma =3$) regression models are -2101.80(2.382, 2.742, 1.151), -2078.23(2.383, 2.175, 0.913), -2077.02(2.383, 2.157, 0.905), -2077.88(2.384, 2.166, 0.909), and -2048.77(2.386, 2.512, 1.053), respectively. The numbers in the parentheses represent an estimated variance, mean, and variance-mean ratio. The SUE ($\gamma =3$) provides the best fit of log-likelihood to the data, although its variance-mean ratio (1.053) disagrees with the actual data (0.977).  It means that the shape of the fertility data distribution resembles the SUE ($\gamma =3$) more than the other models (see Figure 4(a)). Note that the fertility data distribution may be the combination of over-, under-, and equidispersed distributions as described later.

For the takeover bids data, the variance-mean ratio is 2.035/1.738 = 1.171. Therefore, the data present overdispersion. The log-likelihood values of the Poison, the gamma, the Weibull, the CMP, and the SUE ($\gamma =1$) regression models are -184.95(1.737, 2.227, 1.282), -180.37(1.736, 1.710, 0.985), -180.21(1.735, 1.635, 0.943), -180.36(1.738, 1.657, 0.954), and -171.31(1.727, 1.478, 0.856), respectively. The Poisson provides the worst fit in terms of log-likelihood to the data even though it presents overdispersion as the data do. It interprets that the shape of the takeover bids data distribution resembles the Poisson less than the other models (see Figure 4(b)).

\begin{figure}[!t]
	\begin{subfigure}{1\textwidth}
		\centering
		\includegraphics[width=.8\linewidth]{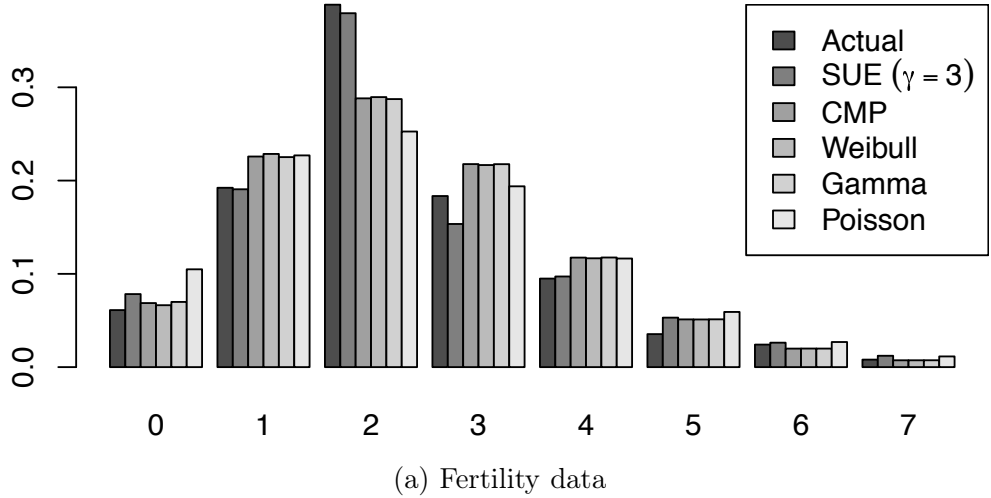}
		\caption{Fertility data}
		\label{fig:sfig1}
	\end{subfigure}%
	
	\begin{subfigure}{1\textwidth}
		\centering
		\includegraphics[width=.8\linewidth]{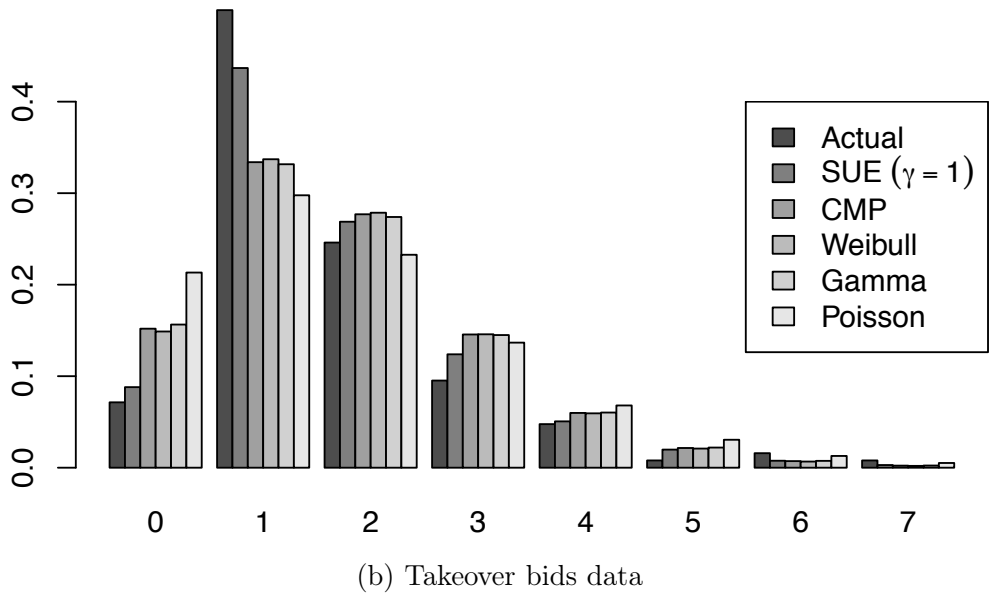}
		\caption{Takeover bids data}
		\label{fig:sfig2}
	\end{subfigure}
	\caption{Sample and predicted relative frequency distributions}
	\label{fig:fig}
\end{figure}

Figure 4 presents the sample probabilities and the predicted probabilities evaluated at individual covariates for the Poisson, the gamma, the Weibull, the CMP, and the SUE. The fertility and the takeover bits datasets contain an excess of two and one outcomes, respectively. It means there are more twos and ones in the two datasets than predicted by the Poisson, the gamma, the Weibull, and the CMP. Figure 4(a) reveals that the models, excluding the SUE ($\gamma =3$), greatly underpredict the two outcomes because the third event is unusual. The SUE ($\gamma =3$) model is assumed the rate between the 2nd and the 3rd event differs from others. Thus, it leads to a considerable improvement of the predicted probabilities in the fertility case. Figure 4(b) shows that all the models underpredict the one outcome because there may be more than one unusual event. However, the SUE ($\gamma =1$) model provides a significant improvement of the predicted probabilities in the takeover bits case, although there is only one unusual event. 

\begin{table}[!t]
	\setlength{\tabcolsep}{0.22em}
	\centering
	\caption{Regression model results for fertility data.}    \
	\makebox[\textwidth]{\begin{tabular}{lrrrrrrrrrr}
			\toprule 
			\multirow{2}{*}{} & \multicolumn{2}{c}{} & \multicolumn{2}{c}{} & \multicolumn{2}{c}{Model} & \multicolumn{2}{c}{} \\
			\cmidrule(l){2-11}
			\multirow{2}{*}{} & \multicolumn{2}{c}{Poisson} & \multicolumn{2}{c}{Gamma} & \multicolumn{2}{c}{Weibull} & 
			\multicolumn{2}{c}{CMP}& \multicolumn{2}{c}{SUE ($\gamma =3$)}\\
			\cmidrule(l){2-3} \cmidrule(l){4-5}\cmidrule(l){6-7} \cmidrule(l){8-9} \cmidrule(l){10-11}
			Variable & \multicolumn{1}{c}{Coef}&\multicolumn{1}{c}{SE}&\multicolumn{1}{c}{Coef}&\multicolumn{1}{c}{SE}&\multicolumn{1}{c}{Coef}&\multicolumn{1}{c}{SE}&\multicolumn{1}{c}{Coef}&\multicolumn{1}{c}{SE}&\multicolumn{1}{c}{Coef}&\multicolumn{1}{c}{SE}\\
			\midrule
			Intercept                   &    1.147 &   0.302   &   1.557   &  0.252  &   1.397    &   0.314  &    1.721   &   0.357 &  1.335  & 0.307 \\
			German                     & -0.200  &  0.072  &  -0.190  &  0.059  &	-0.223   &   0.072  &  -0.266  &   0.084	&	-0.194	 & 0.073\\
			Years of schooling   &   0.034  &  0.032  &   0.032  &  0.027  &	 0.039   &   0.033  &    0.044  &   0.037	&	 0.033  &	0.033\\
			Vocational training   & -0.153  & 0.044   & -0.144   &  0.036  &  -0.173   &   0.044  &  -0.202  &   0.051	&	-0.158	 & 0.044\\
			University                  &-0.155  & 0.159    & -0.146    &  0.130  &   -0.181   &   0.160  &  -0.207  &   0.182	&	-0.136	&  0.162\\
			Catholic                     &   0.218 &  0.071   &  0.206    &  0.058  &   0.242   &   0.070  &   0.289   &   0.082	&	0.212	&  0.071\\
			Protrstant                  &  0.113   & 0.076   &   0.107    &  0.062  &   0.123    &   0.076  &    0.151   &   0.088	&	0.097  & 0.077 \\
			Muslim                       &   0.548 &  0.085  &   0.523   & 0.070  &    0.639   &  0.087  &    0.742   &  0.103  &  0.547 & 0.087 \\
			Rural                          &   0.059  & 0.038    &  0.055   &  0.031  &   0.068    &   0.038  &    0.078 &   0.044 &  0.062 & 0.039 \\
			Year of birth              &   0.002 &  0.002  &  0.002   &  0.002   &   0.002   &   0.002  &    0.003  &  0.003 &  0.001   &0.002\\
			Age at marriage        &  -0.030 &  0.007	  & -0.029  &  0.005  & -0.034   &   0.006  &  -0.040   &  0.008 & -0.030  & 0.007\\
			ln $\alpha$                &                &                 &   0.364   &  0.049 &   0.212   &   0.027   &    0.357   &  0.047 & -0.652 &   0.064\\
			Log likelihood & \multicolumn{2}{c}{-2101.80} & \multicolumn{2}{c}{-2078.23} & \multicolumn{2}{c}{-2077.02} & 
			\multicolumn{2}{c}{-2077.88}& \multicolumn{2}{c}{-2048.77}\\
			Elapsed time (seconds) & \multicolumn{2}{c}{0.01} & \multicolumn{2}{c}{132.29} & \multicolumn{2}{c}{37.61} & \multicolumn{2}{c}{4.78}& \multicolumn{2}{c}{0.63}\\
			\bottomrule
	\end{tabular}}
\end{table}

Tables 1 and 2 present the results from regressions for the number of children and the number of bids data. The regression results from the gamma and the Weibull models are produced by the "nlminb" method and the CMP and the SUE models by the "BFGS" method. The four models use the Poisson coefficients in Tables 1 and 2 as starting values of the unknown parameters $\beta _{0},\beta _{1},\beta _{2},...,\beta _{r}$, and the initial value of $\beta _{r+1}$ (the shape parameter) is set to zero. The Poisson coefficients are perhaps the best initial guess for these models because the models generalize the Poisson. Typically, the different optimization methods and starting points may yield different results. 
For example, the log-likelihood values of the gamma, the Weibull, the CMP, and the SUE ($\gamma =3$) for the number of children computed using the "CG" method whose starting values equal zero are -2081.66, -2078.15, -2124.67, and -2145.72. These values are worse than the values in Table 1. The gamma model uses the "BFGS" method, the Weibull model the "BFGS" method, and the SUE ($\gamma =3$) model the "nlminb" method, which starting points equal the Poisson coefficients. Their log-likelihood values are -2078.25, -2077.03, and 2048.76, respectively. These values are nearly identical to the values in Table 1.

\begin{table}[!t]
	\setlength{\tabcolsep}{0.22em}
	\centering
	\caption{Regression model results for takeover bids data.}    \
	\makebox[\textwidth]{\begin{tabular}{lrrrrrrrrrr}
			
			\toprule 
			\multirow{2}{*}{} & \multicolumn{2}{c}{} & \multicolumn{2}{c}{} & \multicolumn{2}{c}{Model} & \multicolumn{2}{c}{} \\
			\cmidrule(l){2-11}
			\multirow{2}{*}{} & \multicolumn{2}{c}{Poisson} & \multicolumn{2}{c}{Gamma} & \multicolumn{2}{c}{Weibull} & 
			\multicolumn{2}{c}{CMP}& \multicolumn{2}{c}{SUE ($\gamma =1$)}\\
			\cmidrule(l){2-3} \cmidrule(l){4-5}\cmidrule(l){6-7} \cmidrule(l){8-9} \cmidrule(l){10-11}
			Variable & \multicolumn{1}{c}{Coef}&\multicolumn{1}{c}{SE}&\multicolumn{1}{c}{Coef}&\multicolumn{1}{c}{SE}&\multicolumn{1}{c}{Coef}&\multicolumn{1}{c}{SE}&\multicolumn{1}{c}{Coef}&\multicolumn{1}{c}{SE}&\multicolumn{1}{c}{Coef}&\multicolumn{1}{c}{SE}\\
			\midrule
			Intercept                    &   0.986  &  0.534  & 1.609  &  0.432  & 1.330   & 0.548 &   1.832   &  0.720  &  0.653  & 0.568    \\
			Leglrest                     &   0.260  &  0.151   &   0.234 &  0.111  &	 0.318   & 0.153   &  0.391   &  0.191	& 0.345   & 0.162      \\
			Rearest                      & -0.196  &  0.192  & -0.168 &  0.142  &	-0.244  & 0.193   & -0.307  &  0.240	&-0.385  &	0.209      \\
			Finrest                       &   0.074  &  0.217  &  0.072 &  0.160  &  0.043   & 0.217  &  0.113   &  0.268	&	0.016	 & 0.229      \\
			Whtknght                   &  0.481  &  0.159  &  0.430 &  0.117  &  0.568    & 0.162  &  0.710  &   0.210	&	0.664	 &  0.176      \\
			Bidprem                     & -0.678  &  0.377  &-0.616 &  0.278  & -0.791   & 0.381 & -1.009    &   0.477	& -0.876	 &  0.406     \\
			Insthold                     & -0.362  &  0.424  &-0.323 &  0.313  &-0.445   & 0.426  &-0.546  &  0.521	& -0.569  & 0.461      \\
			Size                            &   0.179  &  0.060  &  0.164 &  0.045  &  0.218   & 0.062 &   0.283  &  0.082  &  0.251   & 0.065      \\
			Sizesq                        & -0.008  &  0.003  &-0.007 &  0.002  &-0.010   & 0.003 & -0.012  &  0.004  & -0.011  & 0.003     \\
			Regulatn                     & -0.030 &  0.161  &-0.024  &  0.119  &-0.042   & 0.160  & -0.041  &  0.197   & -0.039  & 0.170       \\
			ln $\alpha$                 &                &                 & 0.544  &  0.161  & 0.331    & 0.093  &  0.551  &  0.152   & 1.086    & 0.218		  \\
			Log likelihood & \multicolumn{2}{c}{-184.95} & \multicolumn{2}{c}{-180.37} & \multicolumn{2}{c}{-180.21} & 
			\multicolumn{2}{c}{-180.36}& \multicolumn{2}{c}{-171.31}\\
			Elapsed time (seconds) & \multicolumn{2}{c}{0.00} & \multicolumn{2}{c}{8.70} & \multicolumn{2}{c}{2.46} & \multicolumn{2}{c}{0.42}& \multicolumn{2}{c}{0.04}\\
			\bottomrule
	\end{tabular}}
\end{table}

Comparing $\alpha$ in Table 1, these values in the gamma, the Weibull, and the CMP regression models are respectively 1.439, 1.236, and 1.429, which exceed one considerably, so there is an indication of underdispersion. These three regression models with fixed $\alpha$ exhibit only one of over-, under-, and equidispersion. In other words, the dispersion types of these regression models depend only on $\alpha$ but not on $\lambda$.
The SUE ($\gamma = 3$, $\alpha = 0.521$) regression model displays overdispersion ($\lambda > 3.67$), underdispersion ($\lambda < 3.67$), and equidispersion ($\lambda = 3.67$) (see Figure 3(c)) because the dispersion types of the SUE ($\gamma > 1$) regression models depend on $\alpha$ and $\lambda$. It shows the flexibility of the SUE ($\gamma > 1$) regression model to allow for over-, under-, and equidispersion, although $\alpha$ is a fixed number. This property does not appear in the gamma, the Weibull, and the CMP count models.

\begin{figure}[!t]
	\begin{subfigure}{0.5\textwidth}
		\centering
		\includegraphics[width=0.9\linewidth]{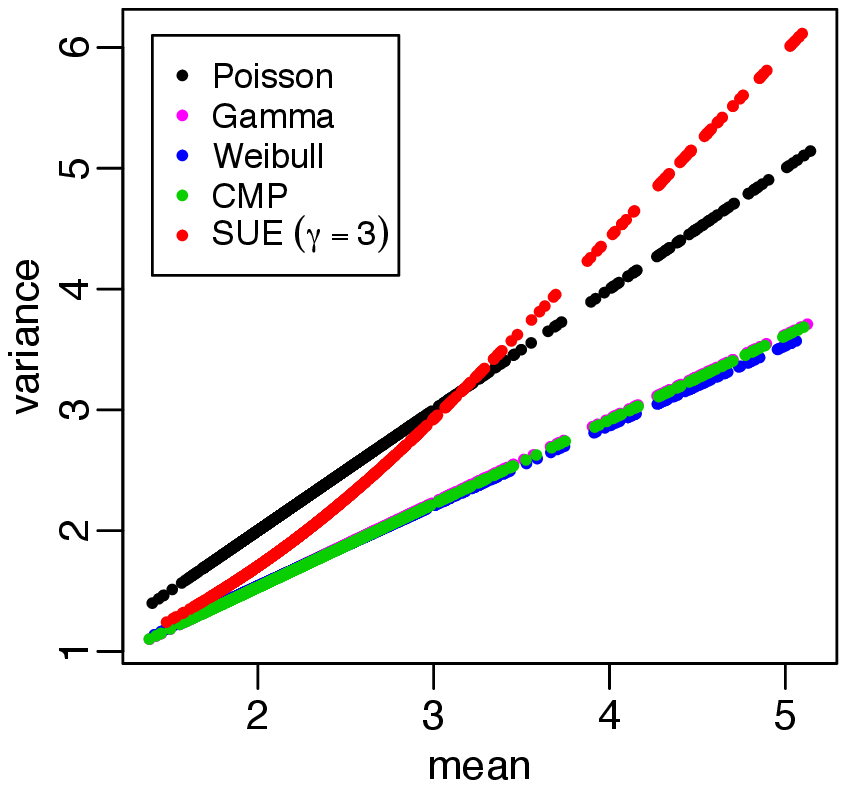}
		\caption{Fertility data}
		\label{fig:sfig1}
	\end{subfigure}%
	\begin{subfigure}{0.5\textwidth}
		\centering
		\includegraphics[width=0.9\linewidth]{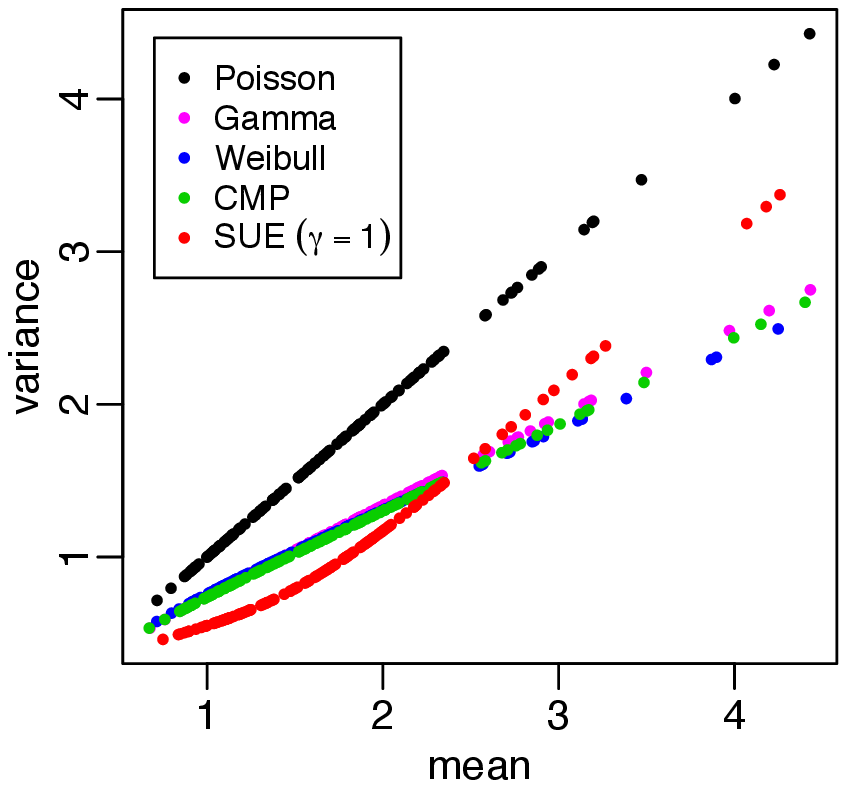}
		\caption{Takeover bids data}
		\label{fig:sfig2}
	\end{subfigure}
	\caption{Scatterplots of estimated variances versus estimated means.}
	\label{fig:fig}
\end{figure}
Figure 5 presents scatterplots of the fertility and the takeover bids data. The dotted points are an ordered pair of the estimated mean and variance of each response variable produced by the five models. The points below and above the 45-degree (Poisson) line indicate underdispersion and overdispersion, respectively. In Figure 5(a), the red curved (SUE ($\gamma =3$)) and the black straight (Poisson) lines cut each other at a point, which is the estimated mean equals the estimated variance. The gamma, the Weibull, and the CMP lines are nearly coincident, indicating a similar ability of these three models to handle the fertility data. It is supported by the results in Table 1 that the log-likelihoods of these models are very similar. According to the SUE ($\gamma =3$) regression model, the fertility data are divided into two sets. The first set consists entirely of the underdispersed response variables, and the overdispersed response variables belong to the second. The first set (1150 members) is about 12 times bigger than the second set (93 members). The gamma, the Weibull, and the CMP models in Figure 5(b) can be interpreted similarly as Figure 5(a), but the SUE ($\gamma =1$) is different. The SUE ($\gamma =1$) curved line is below the Poisson line, indicating the response variables are all underdispersed and convex to the left of the intersection point, which contains almost all the SUE ($\gamma =1$) points (114 out of 126). The curved line from a model, which is stronger convex than the SUE ($\gamma =1$) model, may provide a satisfactory fit to the takeover bids data. This model may need more than one unusual event, and it is interesting in future research.

\section{Conclusion}
The Poisson, the gamma, the Weibull, and the CMP count models are well-known, but their underlying assumption of equal rates limits their use in many econometric applications. The SUE count models, in contrast, are assumed that the rates are unequal, and the distributions of their interarrival times are exponential. One significant advantage of these new count models is the dispersion types defined by the rate and the shape parameters. Hence, the SUE count models can display over-, under-, and equidispersion, although the shape parameter is a fixed number. In other words, the conditional variance and mean of the SUE regression models are not linearly related, allowing for a mixture of the over-, under-, and equidispersed distributions. The SUE regression models are applied to the fertility and the takeover bids data, and they offer significant improvements in log-likelihood compared to the above well-known regression models. For fertility data, the results show that the women's intentions to have a third child, an unusual event, are considerably less than other children. The behavior of these women cannot be captured by the above well-known count models with equal rates. Even though the SUE count models offer significant improvements, future studies could improve the models for better results by adding at least one unusual event or replacing the exponential distribution with a non-negative distribution such as the gamma, the Weibull, etc. 
\bigskip
\begin{center}
	{\large\bf SUPPLEMENTARY MATERIAL}
\end{center}
\begin{description}
\item[\textbf{R} package for analyzing SUE count data:] The \textbf{R} package \textbf{SUEcount} provides a function \texttt{glm.sue()} for fitting SUE regression models, and it also contains the fertility and the takeover bids datasets used as examples in this article. The two \textbf{R} scripts \verb|results_fertility.R| and \verb|results_takeoverbids.R| in the \verb|inst\example| directory show the summary statistics in Tables 1 and 2. (ZIP file)   
\end{description}

\bibliographystyle{Chicago}

\bibliography{mong}
\end{document}